\documentclass[cameraready]{Interspeech}

\title{Mind the Gap: Impact of Synthetic Conversational Data on Multi-Talker ASR and Speaker Diarization}

\author[affiliation={1,2}, orcid=0009-0000-4958-202X, equalcontribution=true]{Alexander}{Polok}
\author[affiliation={3}, orcid=0000-0001-5381-3433, equalcontribution=true]{Ivan}{Medennikov}
\author[affiliation={1}, orcid=0000-0002-8800-0210]{Jan}{Černocký}
\author[affiliation={2}, orcid=0000-0002-5970-8631]{Shinji}{Watanabe}
\author[affiliation={1}, orcid=0000-0002-4951-5908]{\\Lukáš}{Burget}
\author[affiliation={2}, orcid=0000-0002-5358-1844]{Samuele}{Cornell}

\address{
    $^1$ Brno University of Technology, Czechia
    $^2$ Carnegie Mellon University, USA
    $^3$ NVIDIA, USA
}

\email{ipoloka@fit.vut.cz}


\usepackage{comment}
\usepackage{cite}
\usepackage{IEEEtrantools}

\begin{document}
\bstctlcite{IEEEexample:BSTcontrol}

\maketitle

\begin{abstract}
Recent breakthroughs in multi-talker ASR (MT-ASR) and speaker diarization (SD) rely on synthetic data to mitigate the scarcity of large-scale conversational recordings, yet the impact of specific simulation choices remains poorly understood. To mind the gap between simulated mixtures and real-world interactions, we present a study of synthetic data generation for leading MT-ASR (DiCoW) and SD (Sortformer) systems. By introducing FastMSS, a highly efficient open-source simulator, we analyze turn-taking dynamics, source domain, acoustic augmentation, and data mixing strategies. Our findings reveal that optimal simulation recipes are highly task-dependent: increasing speech overlap benefits ASR but degrades diarization. Furthermore, broad source diversity consistently outperforms exact domain matching. Ultimately, synthetic-only training approaches real-data baselines, and combining simulated data with real recordings yields substantial gains over real-only training across both tasks.
\end{abstract}

\section{Introduction}

Multi-talker conversational speech processing is undergoing a rapid transformation, driven largely by the shift from highly specialized pipelines to less data-hungry methods built on pretrained foundation models~\cite{radford2023robust,chen2022wavlm,mms,zhang2023googleusmscalingautomatic}. By leveraging massive amounts of single-speaker or self-supervised data, these foundational backbones can be effectively fine-tuned for complex downstream tasks, including multi-talker automatic speech recognition (MT-ASR)~\cite{adapting_zili,li23o_interspeech}, speaker diarization~\cite{han25_interspeech, park2025sortformer}, or spoken dialog models~\cite{veluri-etal-2024-beyond,defossez2024moshispeechtextfoundationmodel}. 
Despite this architectural progress, a persistent bottleneck remains: the severe scarcity of real, large-scale conversational training data. Real meeting corpora, such as AMI~\cite{Mccowan2005_ami}, NOTSOFAR-1~\cite{vinnikov24_interspeech,abramovski2025summary}, MCoREC~\cite{nguyen2026cocktailpartybenchmarkmultimodaldataset}, and CHiME-5/6 \cite{watanabe2020chime,barker18_interspeech} provide spontaneous conversational data but are limited in scale, typically comprising only tens to a few hundreds of hours. Furthermore, annotating real conversational data is prohibitively expensive, time-consuming, and frequently complicated by privacy concerns. To fully unlock the capabilities of foundational models, the field has increasingly relied on synthetic data generation—ranging from heuristic signal-mixing~\cite{mms_msg,park23b_chime} to text-to-speech (TTS) pipelines~\cite{cornell2024generating, burdisso2025sdialogpythontoolkitendtoend, Voicebox}. Despite this widespread adoption, the community's approach to simulation remains heavily fragmented, leaving three crucial gaps in our understanding: First, existing simulation strategies are almost exclusively designed and evaluated for a single specific task~\cite{landini22_interspeech,broughton25_interspeech,sim_mt_asr,mt_asr_v2, bamfo-odoom-etal-2024-synthetic, hilmes24_syndata4genai, cornell2024generating}, leaving it unclear if the same synthetic data properties benefit complementary multi-talker tasks, such as multi-talker ASR (MT-ASR) and diarization. Second, studies often rely on a single source of seed data~\cite{libricss, Cosentino2020LibriMixAO}, leaving the impact of source domain mismatch and the generalizability to spontaneous, ``in-the-wild'' interactions largely unexplored. 
Finally, it remains unclear to what extent synthetic data can substitute for real conversational recordings, especially for data-hungry tasks such as end-to-end diarization~\cite{eend}. Nor is there consensus on the best strategy for combining the two: training jointly, fine-tuning on real data after synthetic pre-training, or relying on synthetic data alone.



To address these gaps, we present a comprehensive study systematically varying turn-taking dynamics, source domain, acoustic augmentation, and data mixing strategies across two leading models: Sortformer~\cite{park2025sortformer} for diarization and DiCoW~\cite{DiCoW} for MT-ASR.
To enable our experiments and ensure full reproducibility, we release FastMSS,\footnote{\url{https://github.com/popcornell/FastMSS}} a fully open-source conversation simulation toolkit that supports flexible turn-taking modeling and scalable generation of multi-talker long-form synthetic conversations.

Our main findings are: (i) optimal simulation strategies are task-dependent: boosting overlap improves MT-ASR but degrades diarization; (ii) mixing seed utterances from diverse domains outperforms using a single in-domain source, even when the latter is matched to the test set; (iii) acoustic augmentation is critical for diarization but not for MT-ASR; and (iv) synthetic data provides a highly competitive alternative when in-domain recordings are unavailable: for diarization, synthetic-only training performs close to a strong reference model, while combining synthetic and real data yields our best overall performance.



\section{Multi-Talker Speech Processing}
\label{sec:mt_processing}


\subsection{Multi-Talker ASR: DiCoW}

Multi-talker ASR (MT-ASR) has traditionally been tackled through modular separation-based pipelines~\cite{niu24_chime}, end-to-end architectures like Serialized Output Training (SOT)~\cite{kanda2020serialized}, or target-speaker conditioning~\cite{Kanda2019_spkloss}. The latter paradigm has seen rapid advancement with the introduction of diarization conditioning, a data-efficient mechanism for adapting pretrained single-speaker ASR models to overlapping, multi-talker scenarios~\cite{polok2024targetspeakerasrwhisper}. We employ Diarization-Conditioned Whisper (DiCoW)~\cite{DiCoW,polok2026sedicow}, which achieves strong MT-ASR performance on benchmarks like AMI~\cite{Mccowan2005_ami} and LibriMix~\cite{Cosentino2020LibriMixAO} by injecting frame-level diarization cues into a Whisper~\cite{radford2023robust} encoder.

\subsection{Speaker Diarization: Sortformer}

End-to-end neural diarization~\cite{eend} is a promising paradigm thanks to its native handling of overlapping speech and architectural simplicity, but it is inherently data-hungry making it an ideal candidate to study the impact of synthetic data properties.  
In contrast, clustering-based~\cite{pyannote} and hybrid systems~\cite{eend_vc,bredin2021end,plaquet23_interspeech,han25_interspeech} rely on pretrained speaker embeddings and non-differentiable clustering steps, which may obscure the direct effect of synthetic data simulation properties. 

We therefore employ Sortformer~\cite{park2025sortformer,medennikov25_interspeech}, an EEND-based encoder-only model that resolves speaker permutations via Sort Loss in arrival-time order, eliminating attractors~\cite{eend_eda, horiguchi2022online} and clustering entirely. Built on the self-supervised NEST-FastConformer backbone~\cite{nest,fastconformer}, it achieves highly competitive diarization across multiple benchmarks~\cite{park2025sortformer}.

\section{Multi-Speaker Conversation Simulation}\label{sec:toolkit}

To enable controlled and fast experimentation along the axes described above, we developed FastMSS, an open-source multi-speaker conversation simulator focused on scalable generation with native Lhotse~\cite{zelasko2021lhotsespeechdatarepresentation} integration. Given a set of single-speaker utterances from a source dataset, FastMSS generates multi-talker mixtures with configurable turn-taking dynamics and acoustic conditions including room impulse responses via Pyroomacoustics~\cite{pyroomacustics}, support for multiple noise sources and per-utterance gain variation. Word-level alignments, when available, are used to split source utterances at pause boundaries for more realistic turn-taking.

\begin{figure}[t]
      \centering
      \includegraphics[width=0.95\linewidth]{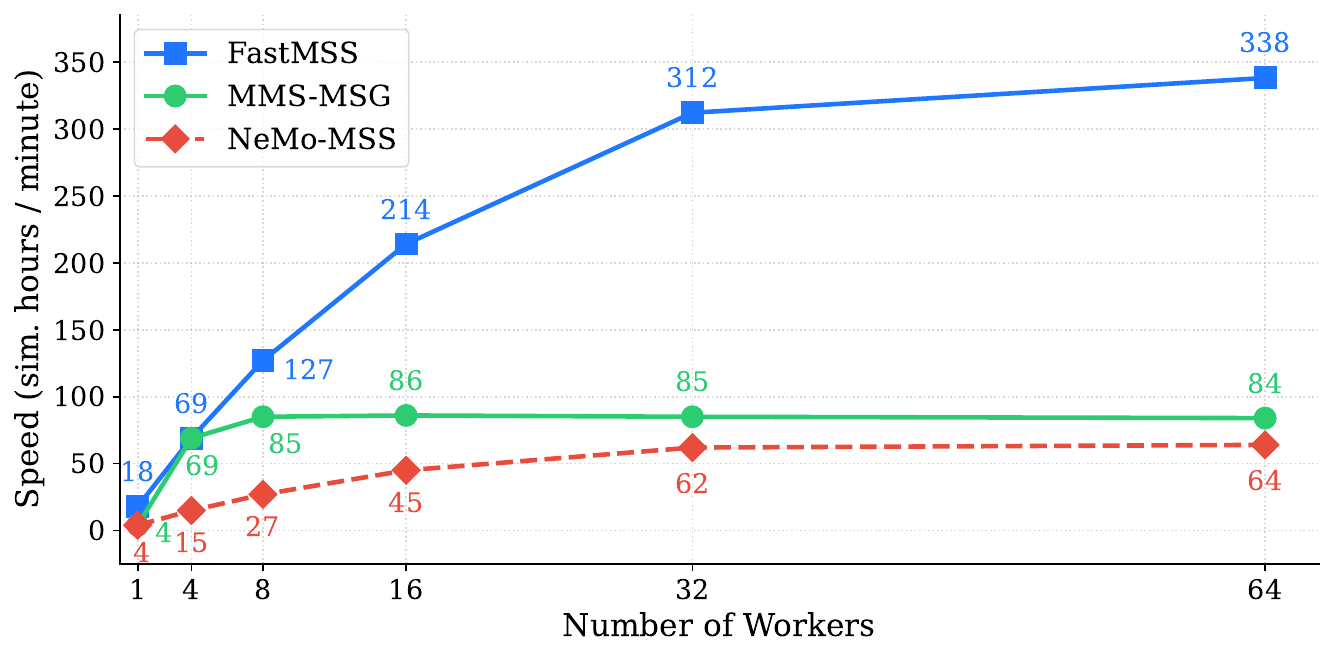}
      \vspace{-0.3cm}
      \caption{Generation speed vs.\ worker count for 6{,}000 two-minute meetings from
      LibriSpeech.}
      \label{fig:scaling}
      \vspace{-0.5cm}
  \end{figure}

Figure~\ref{fig:scaling} demonstrates FastMSS's scalability against two widely used simulators, MMS-MSG~\cite{mms_msg} and the NeMo multi-speaker data simulator~\cite{park23b_chime}, benchmarked on identical hardware (4$\times$
AMD EPYC 7742, 256 CPUs, 1\,TB RAM) generating 6{,}000 two-minute meetings from LibriSpeech \texttt{train-clean-100} without reverberation or noise. FastMSS scales efficiently from single up to 32 processes, generating 1{,}000 hours of annotated multi-talker audio in under five minutes. Instead, MMS-MSG adopts a producer--consumer design in which examples are generated in parallel but audio is written sequentially, creating an I/O bottleneck that caps throughput beyond eight workers. 
NeMo is also process-based, with higher per-conversation overhead from PyTorch tensor operations.

\subsection{Turn-Taking Model}
\label{sec:turn_taking}

FastMSS Turn-Taking (TT) model simply extends the two-speaker HMM-based approach of Yamashita et al.~\cite{yamashita22_odyssey} to an arbitrary number of speakers. Four utterance transition types are modeled: \textit{turn hold} (TH, same speaker continues after a pause), \textit{turn switch} (TS, different speaker after a gap), \textit{interruption} (IR, different speaker with overlap), and \textit{backchannel} (BC, short overlapping utterance from another speaker).

At each step, the transition type is drawn either from a fixed categorical distribution $\mathbf{p} = [p_{\text{TH}}, p_{\text{TS}}, p_{\text{IR}}, p_{\text{BC}}]$ or from a first-order Markov chain with transition matrix $\mathbf{P}$, where $P_{ij} = p(z_t{=}j \mid z_{t-1}{=}i)$. Pause and gap durations (TH, TS) are sampled from exponential distributions with rate parameters $\beta_{\text{TH}}$ and $\beta_{\text{TS}}$; overlap extent (IR) is drawn as a ratio from a truncated exponential with parameter $\beta_{\text{IR}}$; and backchannels are placed uniformly within the preceding utterance span. A constraint ensures no speaker overlaps with themselves. For conversations with more than two speakers, the next speaker is selected uniformly at random from all participants other than the current one (or, for TH, the same speaker continues).
All parameters can be specified manually or fitted automatically from any annotated corpus via maximum likelihood estimation.

A known limitation of concatenative simulation is the lack of inter-turn semantic coherence. While negligible for acoustic diarization, it introduces distribution shifts in TS-ASR. We mitigate this by freezing the ASR decoder during training. Instead of relying on computationally expensive alternatives like TTS generation or perplexity priors, our framework prioritizes scalability and acoustic fidelity over semantic continuity.



\section{Experimental Setup}
\label{sec:setup}

\subsection{Datasets}


As source domains for synthetic generation, we use: LibriSpeech~\cite{librispeech} (read speech, 960h), VoxPopuli~\cite{wang_voxpopuli_2021} (semi-spontaneous parliamentary speech, 543h), otoSpeech~\cite{otoSpeech-full-duplex-processed-141h} (full-duplex conversational speech, 141h), and the close-talk channels of AMI Meeting Corpus~\cite{Mccowan2005_ami} and NOTSOFAR-1 (NSF-1)~\cite{vinnikov24_interspeech} (spontaneous meetings). All datasets were re-aligned using the Montreal Forced Aligner~\cite{mcauliffe17_interspeech} to ensure consistent word-level timestamps. Noises for data augmentation are taken from the MUSAN~\cite{musan2015}, with ``speech'' noises excluded.


For DiCoW, we evaluate primarily on AMI Single Distant Microphone~(SDM) and NSF-1~\cite{vinnikov24_interspeech} Single-Channel~(SC), alongside LibriSpeechMix~\cite{kanda2020joint} (1--3 speaker mixtures) and Mixer6~(MX6)~\cite{Mixer6} CH4 to assess out-of-domain performance.

For Sortformer, we evaluate on NSF-1 SC and Mix of Headset Mics~(MHM), AMI SDM and MHM,  AliMeeting~\cite{Yu2022M2MeT, Yu2022Summary} near and far microphone, DIHARD-III Eval~\cite{ryant21_interspeech} (1--4 speaker subset), and MSDWild~\cite{liu22t_interspeech} (few-talker eval split) to assess multi-domain generalization. As the full-length recordings of AMI and AliMeeting exceed the limits of the offline Sortformer, we evaluate on 180-second chunks for these datasets. Notably, we use forced-alignment based ground-truth labels for AMI, AliMeeting and NSF-1, because default segment-level annotations are not well-suited for diarization due to severe over-segmentation~\cite{horiguchi_asru2025}. 
For real-data comparisons, we use a ${\sim}$314-hour multi-domain training set comprising NSF-1, AMI, AliMeeting, DIHARD-III Dev, and VoxConverse-v0.3~\cite{voxconv}.

\begin{table*}[t]
    \centering
    \caption{Impact of turn-taking dynamics. DiCoW uses 500 hours of simulated conversations generated from the NSF-1 close-talk source (${\sim}$7.5h), Sortformer uses 2,000 hours of conversations simulated from LibriSpeech (${\sim}$960h). For reference, training DiCoW on NSF-1 SC only yields a tcpWER of 19.8 on NSF-1 SC and 25.2 on AMI SDM. Transition parameters $(p_{\text{TH}}, p_{\text{TS}}, p_{\text{IR}}, p_{\text{BC}})$ are: Flat=(0.25, 0.25, 0.25, 0.25); NSF-1=(0.18, 0.22, 0.30, 0.30); CALLHOME=(0.15, 0.21, 0.44, 0.20); CALLHOME(OV)=(0.09, 0.13, 0.54, 0.24).}
    \label{tab:turn_taking}
    
    \setlength{\tabcolsep}{3pt}
    \renewcommand{\arraystretch}{1.05}
    \begin{tabular}{lcc|ccccccccc}
        \toprule
        & \multicolumn{2}{c|}{\textbf{DiCoW (tcpWER$\downarrow$)}} 
        & \multicolumn{9}{c}{\textbf{Sortformer (DER$\downarrow$)}} \\
        \textbf{TT config} & NSF-1 & AMI
        & \multicolumn{2}{c}{NSF-1} 
        & \multicolumn{2}{c}{AMI 180s} 
        & \multicolumn{2}{c}{AliMtg 180s}
        & \multicolumn{1}{c}{DIHARD-III}  
        & \multicolumn{1}{c}{MSDWild} & Macro \\
        & SC & SDM & MHM & SC & MHM & SDM & Near & Far & 1-4spk & Few & Avg. \\
   
        \midrule
 
        Flat prior          &  24.8  &  29.2  &  24.0  &  33.9  &  21.4  &  27.3  &  23.5  &  38.6  &  18.2  &  \textbf{26.7} & 26.7 \\
        NSF-1              &  23.6  &  27.0  &  \textbf{22.1}  &  32.8  &  21.4  &  26.9  &  24.1  &  \textbf{36.7}  &  \textbf{17.0}  &  28.3 & 26.2 \\
        CALLHOME            &  22.8  &  26.3  &  23.6  &  \textbf{32.3}  &  \textbf{20.7}  &  \textbf{26.6}  & \textbf{23.1}  &  36.8  &  17.8  &  27.7 & \textbf{26.1} \\
        CALLHOME (OV boost)&  \textbf{22.1}  &  \textbf{25.1}  &  23.5  &  36.6  &  21.4  &  27.5  &  24.6  &  37.3  &  19.6  &  30.1 & 27.6 \\
        \bottomrule
    \end{tabular}
    \vspace{-0.3cm}
\end{table*}
\subsection{Model Training and Evaluation}
For multi-talker ASR, we train the DiCoW model on a Whisper-large-v3-turbo backbone. To ensure full reproducibility and isolate the impact of our simulation strategy, we strictly adhere to the original SE-DiCoW~\cite{polok2026sedicow} protocol and hyperparameter configurations , all of which we open-source alongside our toolkit.
Performance is evaluated using time-constrained minimum-permutation WER (tcpWER)~\cite{Neumann2023MeetEval} with a 5\,s collar. Consistent with the DiCoW/SE-DiCoW evaluation protocol, we report throughout this paper tcpWER conditioned on ground truth diarization, and we utilize greedy attention-only decoding.

For speaker diarization, we employ an offline 4-speaker Sortformer initialized with a 109M-parameter NEST-L encoder and no extra transformer layers. Diarization experiments are performed using NVIDIA NeMo Framework~\cite{kuchaiev2019nemo} on 60\,s random crops drawn from 90\,s simulated segments, balancing 1--4 speaker sessions at a 1/3/6/10 ratio. Performance is reported as DER with 0\,s collar, averaged across three random seeds.

\section{Results}
\label{sec:results}


\subsection{Impact of Turn-Taking Dynamics}

        

In Table~\ref{tab:turn_taking}, we isolate the effect of turn-taking by varying only the simulator transition model parameters while keeping all other factors fixed. For DiCoW, the source utterances are NSF-1 close-talk (${\sim}$500h simulated from ${\sim}$7.5h); for Sortformer, LibriSpeech (2,000h simulated from 960h), without augmentation.
For both tasks, corpus-fitted turn-taking statistics consistently outperform a flat prior. DiCoW improves from 24.8\% to 23.6\% on NSF-1 when using NSF-1-fitted statistics, while Sortformer drops from 24.0\% to 22.1\% on NSF-1 MHM. CALLHOME-fitted statistics yield a further improvement for DiCoW (22.8\%), suggesting that the faster turn-taking patterns of telephone conversations provide useful training signal for MT-ASR, and also achieve the best macro DER for Sortformer (26.1\% vs.\ 26.7\% for the flat prior).

However, the two tasks \emph{diverge sharply} when overlap is artificially boosted (\emph{OV boost}) beyond natural levels by boosting the probabilities of interruption and backchannel transitions by a factor of 2 and then re-normalizing. For DiCoW, boosting overlap yields the best synthetic-only result (22.1\% on NSF-1, a 2.7\% abs. improvement over the flat baseline), suggesting that exposure to exaggerated overlap encourages more robust target-speaker tracking, consistent with observations in speech separation~\cite{Cosentino2020LibriMixAO}. Remarkably, this configuration matches the performance of a model trained exclusively on real NSF-1 recordings when evaluated on the out-of-domain AMI benchmark (25.1\% vs.\ 25.2\%), demonstrating that well-configured synthetic data can rival real in-domain training on unseen domains.

For Sortformer, the opposite occurs: overlap boosting \emph{degrades} performance substantially, worsening the macro average from 26.1\% to 27.6\%. This suggests that artificially boosted overlap corrupts the learning of precise speaker boundaries, which is the core objective of diarization. 
This finding has a practical implication: a single simulation recipe cannot optimally serve both ASR and diarization. In the experiments that follow, we therefore use the best task-specific configuration: CALLHOME + overlap boosting for DiCoW, and CALLHOME without boosting for Sortformer.

\begin{table}[t]
    \centering
    \caption{Impact of source domain on DiCoW (tcpWER$\downarrow$, \%). All rows use CALLHOME (OV boost) turn-taking without acoustic augmentation (500\,h simulated per source; 2,500\,h for the \emph{Combined} setup). Best synthetic underlined; best overall bold.}
    \label{tab:source_domain}
    \setlength{\tabcolsep}{0.8pt}
    \begin{tabular}{lccccccc}
        \toprule
        \textbf{Source} & NSF-1 & AMI & LS1 & LS2 & LS3 & MX6 & Macro Avg. \\
        \midrule
        LibriSpeech         &  30.3  &  30.7  &  \underline{\textbf{1.7}}  &  2.5  &  4.3  &  14.7 & 14.0  \\
        VoxPopuli           &  34.1  &  35.1  &  2.8  &  4.8  &  8.0  &  21.6 & 17.7  \\
        otoSpeech           &  28.4  &  36.9  &  3.2  &  5.9  &  11.3  &  20.0 & 17.6  \\
        AMI close-talk      &  25.5  &  18.3  &  2.9  &  5.0  &  8.7  &  14.2 & 12.4  \\
        NSF-1 close-talk    &  22.1  &  25.1  &  3.6  &  6.1  &  10.6  &  \underline{13.9} & 13.6  \\
        Combined            &  \underline{20.6}  &  \underline{16.5}  &  1.8  &  \textbf{\underline{2.4}}  &  \textbf{\underline{3.9}}  &  14.7 & \underline{10.0}  \\
        \midrule
        Real (AMI+NSF)      &  17.7  &  15.5  &  2.8  &  5.9  &  10.5  &  12.9 & 10.9  \\
        Real + Combined     &  \textbf{16.3}  &  \textbf{15.2}  &  1.9  &  2.5  &  4.1  &  \textbf{12.7} & \textbf{8.8}  \\
        \bottomrule
    \end{tabular}
    \begin{flushleft}
    \end{flushleft}
    \vspace{-1cm}
\end{table}

\begin{table*}[t]
    \centering
    \caption{Impact of acoustic augmentation (top) and data combination strategies (bottom). For augmentation, both models use LibriSpeech as the source corpus (500\,h simulated for DiCoW; 2,000\,h for Sortformer). For data combination, DiCoW (\emph{Combined}, 2,500\,h) and Sortformer (LibriSpeech, 2,000\,h) utilize noisy, reverberant sources. The Reference corresponds to diar\_sortformer\_4spk-v1\protect\footnotemark[2]~\cite{park2025sortformer} and DiCoW\_v3\_3\protect\footnotemark[3]~\cite{polok2026sedicow} with greedy decoding.}
    \label{tab:aug_and_scaling}
    \setlength{\tabcolsep}{1.5pt}
    \begin{tabular}{@{} l ccccccc | ccccccccc @{}}
        \toprule
        & \multicolumn{7}{c|}{\textbf{DiCoW (tcpWER$\downarrow$)}}
        & \multicolumn{9}{c}{\textbf{Sortformer (DER$\downarrow$)}} \\
        \textbf{Configuration} & NSF-1 & AMI & LS1 & LS2 & LS3 & MX6 & Macro & \multicolumn{2}{c}{NSF-1} 
        & \multicolumn{2}{c}{AMI 180s} 
        & \multicolumn{2}{c}{AliMtg 180s}
        & \multicolumn{1}{c}{DH-III}  
        & \multicolumn{1}{c}{MSD} & Macro \\
        & SC & SDM & -- & -- & -- & CH4 & Avg. & MHM & SC & MHM & SDM & Near & Far & 1-4spk & Few & Avg. \\
        
        \midrule
        \multicolumn{17}{c}{Impact of Acoustic Augmentation} \\
        \midrule
        None (clean)        &  30.3  &  30.7  &  \textbf{1.7}  &  2.5  &  4.3  &  14.7  & 14.0 &  23.6  &  32.3  &  \textbf{20.7}  &  26.6  &  23.1  &  36.8  &  17.8  &  27.7 & 26.1 \\
        + noise             &  28.3   &  31.5  &  \textbf{1.7}     &  2.4  &  4.0  &  \textbf{14.0}  & \textbf{13.7} &  \textbf{19.4}  &  28.7  &  20.8  &  25.9  &  21.8  &  38.1  &  17.4  &  \textbf{24.3} & 24.6 \\
        + rvb            &  30.1   &  \textbf{31.4}  &  1.8  &  2.7    &  5.3   &  14.3  & 14.3 &  22.9  &  30.2  &  21.5  &  24.9  &  22.9  &  25.7  &  17.6  &  28.7 & 24.3 \\
        + noise+rvb         &  \textbf{28.0}  &  32.5  &  \textbf{1.7}  &  \textbf{2.3}  &  \textbf{3.8}  &  14.4  & 13.8 &  20.7  &  \textbf{25.9}  &  22.0  &  \textbf{23.9}  &  \textbf{21.5}  &  \textbf{22.9}  &  \textbf{16.3}  &  \textbf{24.3} & \textbf{22.2} \\        
        
        \midrule
        \multicolumn{17}{c}{Combining Synthetic and Real Data} \\
        \midrule
        
        Synthetic only      &  20.1   &  16.0  &  \textbf{1.8}  &  \textbf{2.4}  &  4.0  &  14.7  & 9.8 &  20.7  &  25.9  &  22.0  &  23.9  &  21.5  &  22.9  &  16.3  &  24.3 & 22.2 \\
        Real only           &  17.7   &  15.5  &  2.8  &  5.9  &  10.5  &  12.9  & 10.9 &  14.8  &  21.5  &  15.0  &  19.9  &  13.5  &  15.6  &  15.5  &  23.5 & 17.4 \\
        Real + synthetic    &  \textbf{16.3}  &  15.2  &  1.9  &  2.5  &  4.1  &  12.7  & 8.8 &  15.2  &  19.7  &  15.2  &  18.3  &  12.6  &  15.2  &  \textbf{14.0}  &  20.5 & 16.3 \\
        Synthetic $\to$ real &  \textbf{16.3} & \textbf{14.9} & 1.9  & 2.5 & \textbf{3.9} & 12.4 & 8.7 &  \textbf{12.7}  &  \textbf{18.3}  &  \textbf{14.5}  &  \textbf{18.0}  &  \textbf{12.0}  &  \textbf{14.7}  &  \textbf{14.0}  &  \textbf{19.9} & \textbf{15.5} \\
        \midrule
        Reference          &  \textbf{16.3}   &  15.1  &  \textbf{1.8}  &  2.5  &  4.0  &  \textbf{11.7}  & \textbf{8.6} &  19.7  &  25.3  &  20.7  &  26.1  &  18.8  &  32.7  &  15.8  &  22.9  & 22.8 \\

        \bottomrule
    \end{tabular}
    \vspace{-0.4cm}
\end{table*}

\subsection{Impact of Source Domain}

In Table~\ref{tab:source_domain}, we fix DiCoW's best turn-taking configuration and vary the source utterances to quantify domain mismatch effects.

Read speech (LibriSpeech) yields strong performance on the matched LibriSpeechMix benchmarks (1.7\% on LS1Mix) and transfers well to Mixer6 (MX6)(14.7\%), but degrades heavily on real meetings (30.3\% on NSF-1).  Semi-spontaneous speech (VoxPopuli) and full-duplex conversational speech 
(otoSpeech) both underperform relative to LibriSpeech on out-of domain benchmarks, likely due to a combination of domain mismatch, accented speech, and noisier forced-alignment labels as parliamentary and conversational speech is inherently harder to align accurately than read speech.
In-domain close-talk sources help substantially (18.3\% on AMI SDM, and 22.1\% on NSF-1) but each source remains domain-biased. The \emph{Combined} mixture of all sources yields the best synthetic-only meeting results (20.6\% on NSF-1, 16.5\% on AMI) while remaining competitive on LSMix and Mixer6.


Notably, the synthetic-only \emph{Combined} setup already outperforms training on real data alone in macro average (10.0\% vs.\ 10.9\%), confirming that source diversity outweighs exact domain matching (AMI and NSF-1 close talk). 
Combining synthetic with real recordings (\emph{Real + Combined}) further 
reduces the macro average to 8.8\%, outperforming real-only training on 
every benchmark (16.3\% vs.\ 17.7\% on NSF-1; 15.2\% vs.\ 15.5\% on AMI; 
12.7\% vs.\ 12.9\% on Mixer6), demonstrating that synthetic 
data acts as an effective complement, not just a substitute. 

\subsection{Impact of Augmentation}
Table~\ref{tab:aug_and_scaling} (top) investigates the effect of augmentation. To isolate augmentation from source domain effects, we use LibriSpeech. Its studio-quality recordings ensure performance changes are attributable to the augmentation rather than domain mismatch.

For DiCoW, augmentation has a \emph{modest} effect overall: adding noise improves the macro average from 14.0\% to 13.7\%, while adding reverberation does not improve alone, combined with noise yields minor gains on NSF-1 and LSMix but no macro improvement (13.8\%). We attribute this to Whisper's large-scale pretraining, which already provides strong robustness to noise and reverberation, making turn-taking dynamics and source domain the dominant factors for MT-ASR.


For Sortformer, augmentation is \emph{crucial}, but the specific type matters. Adding noise alone reduces macro DER from 26.1\% to 24.6\% and NSF-1 SC from 32.3\% to 28.7\%, but \emph{worsens} far-field performance on AliMeeting (36.8\% $\to$ 38.1\%), suggesting that noise without reverberation does not capture the dominant acoustic mismatch in far-field conditions. Adding reverberation yields far more substantial gains, reducing AliMeeting Far from 36.8\% to 25.7\%, an 11 point absolute reduction. Combining noise and reverberation achieves the best overall result, improving macro DER from 26.1\% to 22.2\%, and constitutes the optimal augmentation recipe for diarization. 

\footnotetext[2]{\url{https://huggingface.co/nvidia/diar\_sortformer\_4spk-v1}}
\footnotetext[3]{\url{https://huggingface.co/BUT-FIT/DiCoW\_v3\_3}}
\subsection{Combining Synthetic and Real Data}
Table~\ref{tab:aug_and_scaling} (bottom) summarizes the best prior configurations and examines how synthetic data interacts with real recordings. 

For DiCoW, adding noise and reverberation on top of the \emph{Combined} source setup yields a marginal 0.2\% absolute improvement in macro tcpWER, giving the best synthetic-only result of 9.8\%. Combining this with real data further reduces the macro average to 8.8\%. 
Fine-tuning a synthetically trained DiCoW model on real data (\emph{Synthetic $\to$ real}) provides slight improvements over joint training (\emph{Real + synthetic}), dropping the overall macro average to 8.7\% and pushing performance on AMI to 14.9\%. A notable performance gap compared to the reference model remains only on Mixer6 (a 0.7\% absolute difference), which we attribute to the reference model's use of LibriMix during training. Although LibriMix does not reflect natural turn-taking, its construction---where one speaker is continuously overlapped by another---likely forces the model to learn more robust target-speaker tracking. This observation is consistent with our overlap-boosting findings in Table~\ref{tab:turn_taking}.

For Sortformer, the best synthetic-only setup (LibriSpeech, CALLHOME turn-taking, 
noise+reverb) already achieves a macro DER of 22.2\%, performing close to the 
reference open-source model despite using only read speech as source. 
This is a fair out-of-domain comparison: the reference model was trained 
without access to NSF-1 and AliMeeting, and the gap is particularly large 
on AliMeeting Far (22.9\% vs.\ 32.7\%), where reverberation augmentation 
proves decisive. Combining simulated with real in-domain data 
(\emph{Real + synthetic}) further improves macro DER to 16.3\%, 
outperforming real-data-only training (17.4\%). A two-stage strategy 
(\emph{Synthetic $\to$ real}), first training on synthetic data then 
fine-tuning on real recordings, achieves the best overall result 
(15.5\% macro DER), outperforming joint training by a large margin. These results demonstrate that well-configured 
synthetic data can approach the performance of real recordings on its own, 
and that combining the two yields further substantial gains for 
end-to-end diarization.

\section{Conclusions}
\label{sec:conclusions}
We presented a systematic study of synthetic conversational data for multi-talker speech processing, investigating the impact of turn-taking dynamics, source domain, and data combination strategies on target-speaker ASR (DiCoW) and speaker diarization (Sortformer). Our main findings are fourfold: (i) optimal simulation recipes are task-dependent, as artificially boosting speech overlap improves ASR performance but degrades diarization accuracy; (ii) a diverse mixture of source domains consistently outperforms any single-domain source; (iii) combining noise with reverberation is critical for diarization, yielding a nearly 4\% absolute macro DER reduction, while providing only marginal gains for DiCoW; and (iv) well-crafted synthetic data is a powerful augmentation tool. Specifically, utilizing a two-stage training strategy (pre-training on synthetic data followed by fine-tuning on real data) achieves a substantial improvement over training on real data only. 
We release FastMSS as an open-source toolkit to support reproducible research in multi-talker data simulation.


\section{Acknowledgements}
This work was partially conducted at the 2025 JSALT workshop. Support was provided by the Ministry of Education, Youth and Sports of the Czech Republic (MoE) through the OP JAK project ``Linguistics, Artificial Intelligence and Language and Speech Technologies: from Research to Applications'' (ID:CZ.02.01.01/00/23\_020/0008518), and Brno Ph.D. Talent Scholarship Programme. Computing on IT4I supercomputer was supported by MoE through the e-INFRA CZ (ID:90254).

\section{Generative AI Use Disclosure}
Generative AI tools have only been used to help revise the manuscript.

\bibliographystyle{IEEEtran}
\bibliography{mybib}

\end{document}